# Dynamics of Open-Source Software Developer's Commit Behavior: An Empirical Investigation of Subversion

Yutao Ma[1]    Yang Wu[2]    Youwei Xu[3]

[1]State Key Laboratory of Software Engineering, Wuhan University, Wuhan 430072, China
[2]Information Center, LingYun Science and Technology Group Co., Ltd., Wuhan 430040, China
[3]Information Center, Wuhan Housing Security and Management Bureau, Wuhan 430015, China

ytma@whu.edu.cn

## ABSTRACT
Commit is an important operation of revision control for open-source software (OSS). Recent research has been pursued to explore the statistical laws of such an operation, but few of those papers conduct empirical investigations on commit interval (i.e., the waiting time between two consecutive commits). In this paper, we investigated software developer's collective and individual commit behavior in terms of the distribution of commit intervals, and found that 1) the data sets of project-level commit interval within both the lifecycle and each release of the projects analyzed roughly follow power-law distributions; and 2) lifecycle- and release-level collective commit interval on class files can also be best fitted with power laws. These findings reveal some general (collective) collaborative development patterns of OSS projects, e.g., most of the waiting times between two consecutive commits to a central repository are short, but only a few of them experience a long duration of waiting. Then, the implications of what we found for OSS research were outlined, which could provide an insight into understanding OSS development processes better based on software developers' historical commit behavior.

## Categories and Subject Descriptors
D.2.9 [**Management**]: Life cycle, Programming teams, Software configuration management;
K.6.3 [**Software Management**]: Software development, Software maintenance.

## 1. INTRODUCTION
Prime examples of open-source software (OSS), e.g., Apache HTTP Server, Mozilla Firefox, GNU/Linux operating system and MySQL, have widely been used by various governments, enterprises and individuals. It has been recognized that the open collaborative development among software developers from all over the world is a significant factor to the success of OSS projects [1]. Revision control, also known as version control, is essential for the organization of multi-developer projects. Within the community of software engineering, the main role of revision control is to track and provide control over changes to source code [2]. As we know, open-source software developers often use revision control tools such as CVS (Concurrent Versions System), SVN (Subversion) and Git to manage and maintain different types of files stored in a (code) repository that hosts an OSS project.

Indeed, such a repository is a kind of file server, but its special feature is that as the files in the repository are changed, each version of those files will be recorded so as to constitute a log of development and maintenance process. With the support of these software tools for revision control, software developers can check out a given revision of a file or the most recent files from the repository to their local workspace of integrated development environments, and commit changes to the file(s) with related short comments/messages to the repository. For centralized revision control systems such as SVN, software developers must serialize their work. After a commit (or so-called code contribution) completes, the client of revision control software informs you of the new revision number, and each successive commit increases the revision number by one [3].

The development of an OSS project under centralized revision control is comprised of making code contributions (i.e., commits) to a central repository that hosts the project [4]. In order to understand OSS development and maintenance processes better, a number of papers have been pursued to investigate the statistical features and quantitative analysis of such an important operation. However, previous work [4] [5] [6] focused mainly on the distribution of commit size, which describes the probability that a given commit is of a particular size in terms of the number of files or lines of code (LOC). Although commit size can be used to estimate a software developer's contributions to an OSS project [7], to the best of our knowledge, few of previous papers conducted empirical research on the dynamics of software developer's individual and collective commit behavior.

Actually, the development of an OSS project could be deemed as a collaborative process of software developers' collective commit behavior [8], which is driven by individual software developer's commit behavior. Human behavior, as one of the significant issues in science, has a history of about one century since the time of Watson [9]. In 2005, the publication of Professor Albert-László Barabási's paper [10] highlighted the research in this area, and provided a basic method for exploring the statistical laws of human behavior from the historical records of human actions. Interestingly, in computer science, several kinds of user behavior, e.g., email communication [10], web server login [11] and Linux command logs [12], have also been found to share a heavy-tailed distribution instead of Poisson distribution or normal distribution. However, few researchers have yet investigated the dynamics or patterns of software developer's commit behavior from the perspective of mining OSS repositories [13].

An OSS project is typically created as a collaborative effort in which software developers improve upon source code and share changes within their communities by using open-access code repositories. Compared with the traditional closed source software development process, the basic principles how and why the collective development of OSS works remain unclear [14]. Analyzing software developer's commit behavior has been considered as a feasible way to investigate this problem [4] [5] [15] [19]. Moreover, such analysis on commit behavior can provide us with relevant and empirically validated insights into how we can improve software quality and collective software development efficiency further [15]. On the other hand, a large quantity of data about software developer's commit behavior on the Internet is readily available to researchers in code repositories, in mailing list archives, and on project websites, which would be a sound case of understanding the dynamics of human behavior on a collective scale in software engineering.

As a starting point for estimating the effects of software developer behavior on the development of OSS projects, the main purpose of this paper is to gain a deeper understanding of the dynamics or patterns of software developer's commit behavior in centralized revision control systems. So, in this paper we will analyze four representative projects on the Apache.org in an attempt to answer the following questions: 1) whether software developer's individual and collective commit behavior would follow some universal laws or patterns; and 2) what are the implications of our findings for OSS research and practice. Based on an empirical investigation, we hope our research outcomes could offer new insights into understanding OSS development processes better, as well as novel ideas for schedule planning and resource allocation of the development of an OSS project.

The remainder of this paper is structured as follows. Section 2 introduces related work. Section 3 explains the analysis method we followed, and presents the primary research results. Section 4 discusses the implications of what we found for OSS development and maintenance. Finally, Section 5 concludes this paper and puts forward future work.

## 2. RELATED WORK
### 2.1 Commit Size Distribution

As mentioned above, commit is an important activity for OSS development, and recently there are a growing number of studies on the size of software developer's commits to OSS repositories. In 2008, the commit size in terms of the number of files was found to follow a Pareto distribution [5] by Hattori *et al*. One year later, Arafat *et al*. found that the commit size in terms of source lines of code (SLOC) follows a power-law distribution [4]; similarly, the distribution of the commit size in terms of LOC was confirmed to be best described by a generalized Pareto model [6] by Kolassa *et al*. The distribution of commit size with a long tail indicates that software developers might carry out large-size commits, though they are less likely to occur. In [8], Lin *et al*. defined a new indicator of commit size as the total number of commits per time unit (e.g., one day, one week, or one month), and found that it also follows a power-law distribution. Compared with the previous work, in this paper we will analyze the dynamics of software developer's commit behavior, which focuses on the statistical distribution of commit intervals.

### 2.2 Commit Classification/Categorization

Each commit may have a different intent. For example, some commits fix software bugs, while others provide new functional features. So, the classification or categorization of commits is still vague and unrecognized so far. In order to relate a commit to certain types of activities such as code management and bug fixing, Hattori *et al.* proposed a classification framework in two dimensions, i.e., commit size and the comment of a commit [5]. In [16], according to the version histories of nine OSS projects, the authors tried to characterize a typical commit in terms of the number of files, the number of LOC and the number of hunks committed together, and found that the size categories of commits can be an indicator for the types of maintenance activities being performed. Furthermore, Hindle *et al.* [17] proposed a taxonomy of large-sized commits grouped by their intents, and found that large-sized commits are more perfective while small-sized commits are more corrective. However, such static classifications don't reveal the dynamics or work patterns of software developer's commit behavior.

### 2.3 Validation of User Behavior Dynamics

Based on the increasing evidence from communication to entertainment and work patterns, Barabási *et al.* found that the timing of many human activities within these fields follow non-Poisson statistics, characterized by bursts of rapidly occurring events separated by long periods of inactivity [10]. Interestingly, such heavy-tailed distributions of inter-event times have also been demonstrated in computer science. For example, the time interval between consecutive visits by a selected user to a given website is best approximated with a power-law distribution, in contrast to the exponential expected for Poisson processes [11]. Barabási *et al.* argued that this is a consequence of a decision based queuing process [10], where most of the events with high priority are rapidly executed, while only a few of the events with low priority last very long waiting times. Unfortunately, there are few representative statistical analyses on the dynamics of software developer's individual and collective commit behavior according to those appropriate data sources such as SourceForge, Apache Software Foundation (ASF) and Google Code.

## 3. DATA ANALYSIS AND RESULTS
### 3.1 Metrics for Commit Behavior

**Table 1. Example of revision logs in a SVN repository**

| No. | Modified date | Committer | Message | Files |
|---|---|---|---|---|
| 11231 | 2008-11-07 16:09:35.217 | yegor | copy the ooxml branch to trunk | 2 |
| 11232 | 2008-11-07 17:29:54.133 | josh | improved tasks for fetching jars | 1 |
| 11233 | 2008-11-07 22:17:10.880 | yegor | fixed a typo in the url to junit | 2 |
| 11234 | 2008-11-08 00:57:23.323 | nick | javadocs cleanup | 5 |
| … | … | … | … | … |

For an OSS project under centralized revision control, a revision is a "snapshot" of its repository at a particular moment in time. After a software developer commits local changes to selected files to the repository, the client of revision control software will tell he/she the corresponding revision number. To see an overall picture of what's been happening in the repository, software developers can view the history of revision logs sorted by revision number in ascending order, which essentially reveals software developer's code contributions to the repository in a collaborative manner. Table 1 shows a simple example of revision logs in a SVN repository, and the title of the last column "Files" means the number of files committed.

**Definition 1.** Project-level collective commit interval (PLCCI) is the time difference (or waiting time) between two consecutive revisions (e.g., 11231 and 11232 in Table 1) in a repository. The formal formula of PLCCI is described as follow.

$$\text{PLCCI}_i = time(r_{i+1}) - time(r_i)(i \in N), \quad (1)$$

where $r_i$ means the $i^{th}$ "global" revision after a successful commit and the function $time()$ returns the standard time of each revision.

**Definition 2.** Project-level individual commit interval (PLICI) is the time difference between two adjacent revisions committed by the same committer (e.g., 11231 and 11233 in Table 1). Supposing $R = \{r_i\} = \bigcup_{k=1}^{M} R_k$ and $R_k = \{r_i^k\}$, where $M$ is the number of committers, the formal formula of PLICI is described as follow.

$$\text{PLICI}_i^k = time(g(r_{i+1}^k)) - time(g(r_i^k))(1 \leq k \leq M), \quad (2)$$

where $r_i^k$ means the $i^{th}$ "private" revision committed by the $k^{th}$ committer and $g$ is a conversion function $g: r_i^k \to r_j$.

**Table 2. Example of revision logs of a file**

| Name | Version | Revision No. | Modified Date | Committer |
|---|---|---|---|---|
| XSSFCell.java | 35 | 11231 | 2008-11-07 16:09:35.217 | yegor |
| | 36 | 11232 | 2008-11-07 17:29:54.133 | josh |
| | 37 | 11233 | 2008-11-07 22:17:10.880 | yegor |
| | … | … | … | … |

In fact, a committer once commits one or more changes to files to the SVN repository of an OSS project. For a file stored in the repository, a revision is basically a file that is modified when compared to the previous version. The history of file revision is all the information collected about a given file as it changes over time. Software developers can select a file in the history view of the client of revision control software, and it will show you all versions and branch tags that are associated with the file. Such information would be valuable for the study of file evolution, bug fixing and code refactorying. Table 2 shows a simple example of revision logs for a frequently-modified file listed in Table 1.

**Definition 3.** File-level collective commit interval (FLCCI) is the time difference between two consecutive versions (e.g. 36 and 37 in Table 2) in the revision history of a file. The formal formula of FLCCI is described as follow.

$$\text{FLCCI}_i = time(h(fr_{i+1})) - time(h(fr_i)), \quad (3)$$

where $fr_i$ represents the $i^{th}$ "private" version of a file $f$ and $h$ is a conversion function $h: fr_i \to r_j$.

**Definition 4.** File-level individual commit interval (FLICI) is the time difference between two adjacent file versions committed by the same committer (e.g. 35 and 37 in Table 2). Supposing $FR = \{fr_i\} = \bigcup_{k=1}^{S} FR_k$ and $FR_k = \{fr_i^k\}$, where $S$ is the number of committers who committed at least 2 revisions of the file in question, the formal formula of FLICI is described as follow.

$$\text{FLICI}_i^k = time(p(fr_{i+1}^k)) - time(p(fr_i^k))(1 \leq k \leq S), \quad (4)$$

where $fr_i^k$ means the $i^{th}$ "private" version of a file $f$ committed by the $k^{th}$ committer and $p$ is a conversion function $p: fr_i^k \to r_j$.

### 3.2 Research Issues

*Q1: Are there any general laws for the distributions of PLCCI PLICI, FLCCI and FLICI in the development process of an OSS project?* That is to say, we want to know whether the distributions of commit intervals in terms of the above-mentioned indicators follow some universal forms of fitting functions. If we do find such laws, what is the implication of the findings for OSS research and practice?

As we know, an OSS always tends to evolve through successive releases in an incremental development manner. After a new release is delivered, in order to meet users' new requirements or improve software quality, software developers update the release by adding new functionality, removing redundant components, and changing the existing ones. Hence, changes to specific files committed by software developers become an integral part integrated into future releases [8]. Until now, iterative and incremental development as a modern development practice has widely been recognized in software engineering. Investigating the dynamics of software developer's commit behavior within different phases/stages that generate new releases would be helpful in schedule planning and human resources allocation of an OSS project. Unfortunately, there are rarely previous studies taking stage or phase into account when analyzing the dynamics of human behavior in electronic communication, entertainment, finance, computer science, etc.

*Q2: Do the distributions of these four indicators recur within different stages/phases that create new releases of an OSS project?* In other words, we are especially concerned whether the distributions of commit intervals in terms of our indicators exhibit similarity in an incremental development process. If so, what is the implication for OSS project development and maintenance?

### 3.3 Data Collection

Our analysis method is based on case studies, so we selected four OSS projects written in Java on the Apache.org: Apache POI (http://poi.apache.org/), Tomcat (http://tomcat.apache.org/), Struts2 (http://struts.apache.org/development/2.x/), and Derby (http://db.apache.org/derby/).

The purpose of Apache POI is to create and maintain Java APIs (Application Programming Interfaces) for manipulating various

file formats based upon the Office Open XML standards and Microsoft's OLE (Object Linking and Embedding) 2 Compound Document format. Tomcat is an open source web server and servlet container developed by the ASF, which provides a "pure Java" HTTP web server environment for Java code to run in. Struts2 is an elegant, extensible framework for creating enterprise-ready Java web applications, and it uses and extends the Java Servlet API to encourage developers to adopt the MVC (model–view–controller) architecture. Derby is a lightweight rational database management system (RDBMS) developed by the ASF, which can be embedded in Java programs and used for online transaction processing.

Table 3. Brief introduction to the projects analyzed

| Project | Description | Release | Class | Commit | Committer |
|---|---|---|---|---|---|
| POI | APIs for file processing | 11 | 2,438 | 8,588 | 10 |
| Tomcat | Servlet container | 10 | 1,980 | 14,481 | 16 |
| Struts2 | Framework for web apps | 20 | 1,521 | 9,999 | 27 |
| Derby | RDBMS | 14 | 2,974 | 21,529 | 35 |

These four projects from different application domains were chosen to be experimental subjects based upon that each project under discussion has been active for at least 3 years and attracts over 10 fixed software developers or committers to participate in. Table 3 shows a brief introduction to the projects in question, including the number of releases, the number of class files, the total number of commits analyzed, and the number of committers. For each project, as of October 12, 2012, we retrieved commit history (from the main truck of its SVN repository) and release history (from the website of the project) till the date by using Subclipse (http://subclipse.tigris.org/), and built a commit dataset for each stage (i.e., the duration between two adjacent releases) according to release history.

## 3.4 Data Processing

Popular functions such as power function, exponential function, polynomial function and logarithmic function were utilized to fit release-level (i.e., a phase/stage that generates a new release) and lifecycle-level data sets, so as to indentify the best fitting curve and its corresponding function expression. Because the projects in question are still evolving, we have to use a long period of development time (over three years) to approximate the lifecycle of each project.

Note that we used a cumulative distribution function (CDF) to reduce noise levels during the estimation of the scaling exponent of power function with the method introduced in [18]. In statistics a distribution can be represented as a probability distribution function (PDF), and the CDF in this paper was computed by integrating the PDF, which describes the probability that a commit interval is of a certain length. Hence, it represents the frequency-of-occurrence of $m$ (i.e., commit interval) with value greater than or equal to a given number,

$$P_{\geq}(m) = \sum_{m' \geq m} P(m') \approx \int P(m')d\,m' \sim m^{-r+1} \; (P(m) \sim m^{-r}). \quad (5)$$

## 3.5 Experimental Results

According to experimental data, we examined the dynamics of commit behavior in terms of the distributions of PLCCI, PLICI, FLCCI and FLICI. The primary findings are described as follows.

### 3.5.1 Dynamics of Collective Commit Behavior

The scattered points of lifecycle-level PLCCI in hours and in days are presented in Figure 1, where X axis denotes the length of a commit interval and Y axis represents the number of commit intervals whose lengths are greater than or equal to a given number. It is obvious from the log-log plot that the data of PLCCI at different time scales roughly follow power-law distributions, suggesting that most of the waiting time between two consecutive commits to the SVN repositories of the projects analyzed last several hours (i.e., more than 80% of commit intervals are less than one day), but only a few of them experience a long duration of waiting that exceeds one week.

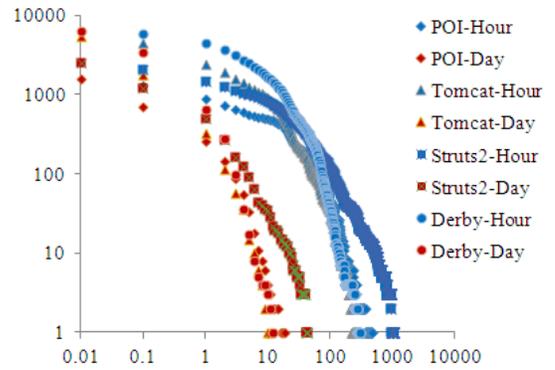

**Figure 1. Distributions of lifecycle-level PLCCI**

On the other hand, interestingly, for the projects analyzed all release-level data about PLCCI in hours can be also best fitted by power functions, implying a general pattern of collectively collaborative commit behavior recurred within different stages in the development process of an OSS project. Due to the space limitations, in this paper we only give an example of POI to illustrate how the data of lifecycle- and release-level PLCCI in hours were fitted (see Table 4, where $R^2$ is the goodness of fit).

After a class file was created in the SVN repository of a project, various software developers would modify it together and then commit changes to it to the repository. The number of revisions to class files has recently been found to follow a power-law distribution [8], implying that most of classes are modified several times, whereas the revisions to a small number of classes are very large. In this paper, for each project under discussion we found that the scattered points of lifecycle-level FLCCI at different time scales roughly follow power-law distributions; similarly, 55 release-level data sets of FLCCI in days were also found to be best fitted by the similar law. The finding indicates that the commit intervals of frequently-modified classes are relatively short on average (i.e., more than 80% of commit intervals of these classes are less than 3 days), while only a minority of classes' revisions need to wait for a very long time of years (see Table 5), perhaps because they are trivial, inactive, deleted, or found to have hidden bugs that are not always easy to detect.

**Table 4. Example of fitting functions for lifecycle- and release-level PLCCI in Apache POI**

| Sample | Logarithmic | $R^2$ | Polynomial | $R^2$ | Exponential | $R^2$ | Power | $R^2$ |
|---|---|---|---|---|---|---|---|---|
| lifecycle | $y=-1.506x+398.570$ | 0.296 | $y=0.015x^2-7.351x+779.360$ | 0.580 | $y=414.43e^{-0.018x}$ | 0.941 | $y=2E+06x^{-2.273}$ | **0.964** |
| 3.5-$\beta$5 | $y=-0.366x+37.311$ | 0.757 | $y=0.005x^2-0.996x+49.701$ | 0.909 | $y=50.849e^{-0.027x}$ | 0.984 | $y=783.07x^{-1.125}$ | **0.996** |
| 3.5-$\beta$6 | $y=-0.183x+35.674$ | 0.485 | $y=0.002x^2-0.743x+60.342$ | 0.791 | $y=37.623e^{-0.017x}$ | 0.882 | $y=4874.9x^{-1.554}$ | **0.992** |
| 3.5-final | $y=-0.184x+29.550$ | 0.652 | $y=0.002x^2-0.591x+42.693$ | 0.865 | $y=37.143e^{-0.017x}$ | 0.969 | $y=755.55x^{-1.082}$ | **0.994** |
| 3.6 | $y=-0.364x+36.344$ | 0.550 | $y=0.008x^2-1.304x+55.159$ | 0.799 | $y=39.381e^{-0.024x}$ | 0.914 | $y=423.64x^{-1.003}$ | **0.973** |
| 3.7-$\beta$1 | $y=-0.079x+28.038$ | 0.380 | $y=0.001x^2-0.341x+48.564$ | 0.654 | $y=26.011e^{-0.008x}$ | 0.878 | $y=2175.8x^{-1.232}$ | **0.974** |
| 3.7-$\beta$2 | $y=-0.103x+14.430$ | 0.614 | $y=0.001x^2-0.340x+20.994$ | 0.833 | $y=16.771e^{-0.018x}$ | 0.908 | $y=256.05x^{-1.017}$ | **0.979** |
| 3.7-$\beta$3 | $y=-0.068x+12.583$ | 0.402 | $y=0.001x^2-0.275x+20.531$ | 0.649 | $y=11.471e^{-0.013x}$ | 0.810 | $y=240.97x^{-1.015}$ | **0.963** |
| 3.7 | $y=-0.148x+11.854$ | 0.797 | $y=0.003x^2-0.382x+15.486$ | 0.934 | $y=15.91e^{-0.034x}$ | 0.952 | $y=148.01x^{-1.071}$ | **0.971** |
| 3.8-$\beta$1 | $y=-0.324x+43.665$ | 0.608 | $y=0.004x^2-1.026x+62.249$ | 0.799 | $y=57.774e^{-0.021x}$ | 0.887 | $y=953.67x^{-1.081}$ | **0.967** |
| 3.8-$\beta$2 | $y=-0.107x+11.531$ | 0.714 | $y=0.001x^2-0.232x+13.552$ | 0.781 | $y=12.775e^{-0.017x}$ | 0.941 | $y=50.21x^{-0.602}$ | **0.949** |
| 3.8-$\beta$3 | $y=-0.092x+16.258$ | 0.521 | $y=0.001x^2-0.373x+27.002$ | 0.842 | $y=14.905e^{-0.016x}$ | 0.791 | $y=584.8x^{-1.226}$ | **0.962** |

Like the analysis on the distribution of PLCCI, Table 5 shows the key statistical results of FLCCI for all class files (.java) of the projects analyzed and the corresponding fitting functions with the best goodness of fit for 55 releases. The parameter $r$ in the last row of Table 5 means power exponent in formula (5) without a unit of measurement.

**Table 5. Key statistical results of FLCCI in days for class files**

| Sample | Mean | Quartile | | | Percentile | |
|---|---|---|---|---|---|---|
| | | 1st | 2nd | 3rd | 90th | 95th |
| Tomcat | 52.88 | 0.04 | 6.99 | 60.80 | 173.59 | 260.85 |
| Struts2 | 81.36 | 0.01 | 10.09 | 61.59 | 242.01 | 477.46 |
| Derby | 128.48 | 1.76 | 23.09 | 145.19 | 286.28 | 623.44 |
| POI | 62.47 | 0.01 | 3.19 | 79.96 | 198.36 | 318.12 |
| release-$r$ | 2.516 | 2.149 | 2.385 | 2.591 | 2.723 | 2.905 |

### 3.5.2 Dynamics of Individual Commit Behavior

Compared with the indicator PLCCI, PLICI indicates the degree of participation of a particular committer to a given project. The shorter a committer's (average) commit interval in a period of time is, the more frequently he/she participates in the project. For all of the projects analyzed, both lifecycle-level and release-level PLICI data sets (i.e., the combined commit intervals of all committers) were found to approximately follow power-law distributions. The finding implies that some of committers work regularly on these projects within short commit intervals (e.g., one day) across various stages in the development process, but the very long waiting times (e.g., several months) between two adjacent commits of the same committer do exist, perhaps because they lose interest in the projects, go on vacation, or complete their tasks and wait for new tasks.

Surprisingly, the dynamics of active committers' individual commit behavior are very similar within both the lifecycle of a project and stages in its development process, though committers have different tasks, backgrounds, habits and expertise. Whether a committer is active is determined by the number of his/her commits in a given period of time. Moreover, we found that the statistical difference of commit intervals between active and inactive committers is very significant.

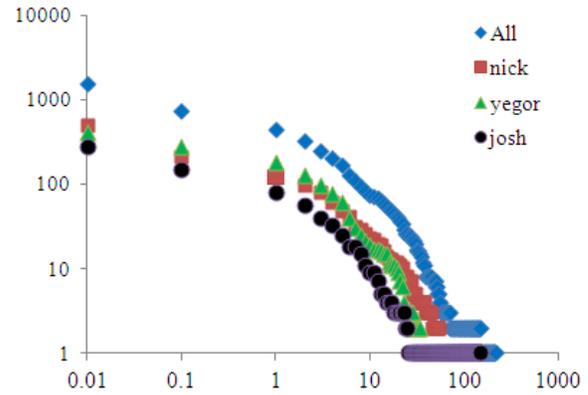

**Figure 2. Distributions of lifecycle-level PLICI**

For example, the distributions of PLICI in days of POI are displayed in Figure 2, where nick, yegor and josh are the top 3 active committers of the project. It is obvious from Figure 2 that they possess a similar behavioral pattern of code commit. Furthermore, Figure 3 shows a box plot for two groups of committers (in POI) labeled as "Active" and "Inactive", where X axis indicates commit interval in days. The plot is interpreted as follows: the left and right of the box are the 25th and 75th percentile (the lower and upper quartiles, respectively), and the red line within the box is the 50th percentile (the median); the black dotted line is the maximum; $p$-value attached to the plot expresses the probability that the observed difference in commit interval between 2 groups of committers is expected by chance, which was calculated by using the Kruskal-Wallis test. Such statistically significant difference illustrates the fact that inactive committers' commit intervals are apt to be irregular and long. We guess this may cause the heavy tail of power-law distributions of lifecycle- and release-level PLICI for all committers.

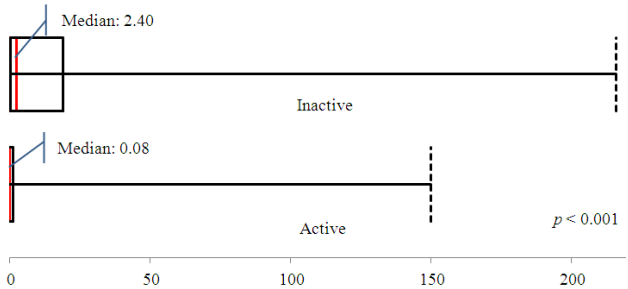

**Figure 3. Box plot for active and inactive committers**

As we expected, lifecycle-level FLICI data sets of the projects in question were also found to share similar heavy-tailed distributions. For frequently-modified class files, on average, the median of commit intervals conducted by the same active committer on the same file is less than 7 days. This implies that some of important classes can be revised and updated by active committers in time. However, such heavy-tailed distributions of FLICI don't recur within each release of these projects, which is mainly due to short durations of some of these releases. In such a short period of time, a tiny minority of committers commit a small number of revisions to only a few files in the SVN repository, so it is difficult to find general laws for the distribution of FLICI.

## 4. DISCUSSION
### 4.1 Implications for OSS Research
*4.1.1 Project-level Commit Behavior*
Although the projects analyzed differ in class size and the number of committers, each project varies slightly in terms of the distribution of PLCCI. Such an indicator reflects to some extent the level of activity of a project developed by various software developers. We argue that power-law distributions of lifecycle- and release-level PLCCI may be derived from the mode of centralized revision control as well as the basic principles of incremental development.

On one hand, SVN uses a centralized model where all the revision control functions take place on a shared server [2], and it stores the latest version of each file in a central repository, with backward-looking differences between two adjacent revisions [3]. So, with the help of SVN client software developers can check out the most current versions of selected files from a SVN repository, and commit their changes to these reference files to the repository as soon as possible. This implies that updates to HEAD of the trunk committed by different committers are always finished quickly, so as to ensure the normal collaborative development among different software developers. So, that is why most of commit intervals in terms of PLCCI are short (see Figure 1).

On the other hand, OSS projects at large follow the principles of incremental development. That is to say, these projects improve their system functions and software quality by delivering new releases one by one. In our empirical experiment, we found an interesting phenomenon that the waiting time between the last commit within the previous release and the first commit after the delivery of a new release is longer than normal values; moreover, it recurs within all 55 releases of the projects in question. Besides accidental events (e.g., the server that hosts the SVN repository is down), this phenomenon (i.e., periodical breaks for making preparations) could be used to explain the occurrence of small probability events (with long waiting time) in power-law distribution of PLCCI.

As mentioned before, collective commit behavior is driven by individual commit behavior. PLICI is an indicator of the degree of participation of particular committers to a given project, which is actually determined by active committers. According to the comparison in the sub-subsection 3.5.3 (see the example of POI in Figure 3), we argue that inactive committers' irregular and long commit intervals may result in the heavy tail of power-law distribution of PLICI for all committers. Although committers differ from each other, the dynamics of active committers' commit behavior are very similar (see the example of POI in Figure 2), implying that they do tend to commit changes to the SVN repository in a short period of time. More interestingly, we found most of active committers like to work regularly on their projects from evening to early morning. This accounts for why the projects analyzed were active during the period.

On the other hand, PLICI for individual active committer roughly follows similar heavy-tailed distribution, where long waiting times between commits may be largely due to personal reasons such as illness, vacation, and accidental events. Considering the importance of committers to a project, how to assign update tasks to remaining committers after one or more active committers temporarily leave the project? It would be an interesting problem for OSS development and practice.

*4.1.2 File-level Commit Behavior*
As we know, division as well as cooperation is the basic principle of modern software development. In a loosely-organized OSS project, software developers share tasks and knowledge resources with social networks tools such as BBS and micro-blogging [20]. When a committer has/receives multiple changes to be committed, he/she has to assess and prioritize the changes to different class files, and then allocates time for the chosen files with high perceived priority. That is to say, committers often rapidly commit changes to those class files with high priority, e.g., user interface, logic control and data processing, and postpone low-priority changes such as adding code comments until the completion of those commits, because too frequent and massive changes committed to the SVN repository are not beneficial to the stability of a project. So, power-law distributions of lifecycle- and release-level FLCCI may be a consequence of a queuing process driven by human decision making based on priority.

In fact, the priority of a class can be simply estimated from the perspectives of functional and structural importance [21]. Within the community of software engineering, functional importance can be measured in terms of WMC (Weighted Method per Class) [22] and SLOC, while structural importance can be assessed in terms of the number of classes that the class in question depends on. Table 6 gives an example of POI to display the correlations between revisions to a class and its functional and structural importance. TOP *k*% in the first column of Table 6 means the top *k*% classes sorted by the number of revisions in descending order. Coefficients for different correlations were calculated in terms of Pearson correlation coefficient. For a 2-tailed test, (*) and (**) indicate that correlations are significant at the 0.95 level and the 0.99 level, respectively.

**Table 6. Pearson Analysis on Correlations**

| Class | Revisions-SLOC | Revisions-WMC | Revisions-Importing Classes |
|---|---|---|---|
| TOP 5% | 0.702** | 0.689** | 0.576** |
| TOP 10% | 0.697** | 0.677** | 0.551** |
| TOP 20% | 0.689** | 0.665** | 0.583** |
| All | 0.546** | 0.538** | 0.552** |

The result of Table 6 indicates that there is a strong positive correlation between importance and revisions, suggesting that important class files do tend to be frequently revised. What we found accords with some of previous work [23] [24]. Based on the above discussion, we argue the finding that the distribution of FLCCI follows a power law may root in the uneven distribution of function and structure among classes. That is to say, committers often give priority to the changes to those classes with complex function and structure, and commit them to the SVN repository in short order. In contrast to the classes with high priority, the waiting times for two consecutive commits of inactive or trivial classes are very long. So, this finding could be a good beginning for investigating new methods for bug detection and program refactoring based on committers' historical commit behavior.

### 4.2 Threats to Validity

SVN has been deemed as a kind of typical centralized revision control software. Distributed revision control systems such as Git take a peer-to-peer approach to version control, as opposed to the client-server approach of centralized systems. Because their work models are different, we are not sure the findings on SVN are still suitable for decentralized revision control systems.

Our experiments were conducted based on the assumption that a software developer modifies the source code of several class files and commits the changes to these files to a central repository. However, in practice this is not always the case. A committer may be an original software developer, be appointed by one of the original developers, or be successfully voted in by the community of committers [25]. According to the information mined from a SVN repository, it is hard for us to distinguish a committer from a software developer. That is to say, if a committer isn't actually a developer, he/she may not commit changes to the SVN repository quickly like a developer in that committers have to obey some commit strategies such as feature freeze [19]. In this paper we focus on the interval between two consecutive commits, and we argue that the problem doesn't affect the overall distribution form we found.

Because some of releases in all experimental releases are of short duration, we didn't find power-law distributions of FLICI within these releases. But in fact, a beta release (see the example in Table 4) generally has many more bugs and performance issues in it than completed software, and it will be kept updating with a new one in a short time till the final version is released. So, the choice of the kind of releases may affect the distribution of release-level FLICI. For example, we guess FLICI data between two adjacent official releases (e.g. 3.6 and 3.7 in Table 4) may roughly follow a power-law distribution.

## 5. CONCLUSIONS AND FUTURE WORK

Revision control software such as SVN has been deemed as an important factor to the success of OSS collaborative development. A commit is the smallest piece of increment a software developer contributes to the SVN repository that hosts an OSS project [19]. Although a few researchers recently began to focus on commit size distribution, commit classification/categorization and developer's contribution estimation, as far as we know, few of those papers conduct empirical studies on the dynamics of software developer's individual and collective commit behavior in terms of commit interval. So, in this paper we took four OSS projects on the Apache.org for example to investigate the general statistical laws for the distributions of lifecycle- and release-level commit intervals. The primary findings are described as follows.

(1) Both lifecycle- and release-level PLCCI roughly follow power-law distributions, suggesting that most of the waiting time between two consecutive commits to a SVN repository are short, but only a few of them experience a long duration of waiting. Such an interesting finding may be derived from the mode of centralized revision control as well as the basic principles of incremental development such as serialized releases.

(2) The distributions of both lifecycle- and release-level PLICI (for all committers) can also be best described by a power-law model. More interestingly, the dynamics of active committers' commit behavior are very similar, and most of them like to work regularly on their projects from evening to early morning. Long waiting times between adjacent commits in the distribution of PLICI for individual active committer may be largely due to personal reasons such as illness, vacation, and accidental events.

(3) The distributions of both lifecycle- and release-level FLCCI are found to share similar laws. We argue that it may be a consequence of a queuing process driven by human decision making based on priority, which is determined by the uneven distribution of function and structure among classes. That is to say, committers often commit the changes to those classes with complex function and structure in short order, while the waiting times for two commits of inactive or trivial classes are longer.

What we found may be interesting collective and individual behavior patterns that software developers work upon an OSS project. Hence, the future work is to design new algorithms for bug detection and program refactoring based on software developers' historical commit behavior, which could help software developers or code reviewers thoroughly verify commits that are more likely to be buggy [26].

## 6. ACKNOWLEDGMENTS

This work is supported by the National Basic Research Program of China under Grant No. 2014CB340401 and the National Natural Science Foundation of China under Grant Nos. 61272111 and 61273216.

## 7. REFERENCES

[1] Crowston, K., Wei, K., Howison, J. and Wiggins, A. Free/Libre Open Source Software Development: What We Know and What We Do Not Know. *ACM Computing Surveys*, 44, 2 (Feb. 2012), Article No. 7.

[2] O'Sullivan, B. Making sense of revision-control systems. *Communications of the ACM*, 52, 9 (Sep. 2009), 56-62.

[3] Collins-Sussman B. The subversion project: buiding a better CVS. *Linux Journal*, 2002, 94 (Feb. 2002), Article No. 3.

[4] Arafat, O. and Riehle, D. The Commit Size Distribution of Open Source Software. In *Proceedings of the 42nd Hawaii International Conference on Systems Science (HICSS'09)* (Hawaii, USA, January 5-8, 2009). IEEE Computer Society Press, New York, NY, 2009, 1-8.

[5] Hattori, L. and Lanza, M. On the nature of commits. In *Proceedings of the 4th International ERCIM Workshop on Software Evolution and Evolvability (EVOL'08)* (L'Aquila, Italy, September 15-16, 2008). IEEE Computer Society Press, New York, NY, 2008, 63-71.

[6] Kolassa, C., Riehle, D. and Salim, M. A Model of the Commit Size Distribution of Open Source. In *Proceedings of the 39th International Conference on Current Trends in Theory and Practice of Computer Science (SOFSEM'13)* (Špindlerův Mlýn, Czech Republic, January 26-31, 2013). Springer-Verlag, Heidelberg, Baden-Württemberg, 2013, 52-66.

[7] Gousios, G., Kalliamvakou, E. and Spinellis, D. Measuring developer contribution from software repository data. In *Proceedings of the 2008 International Working Conference on Mining Software Repositories (MSR'08)* (Leipzig, Germany, May 10-11, 2008). ACM Press, New York, NY, 2008, 129-132.

[8] Lin, S., Ma, Y. and Chen, J. Empirical Evidence on Developer's Commit Activity for Open-Source Software Projects. In *Proceedings of the 25th International Conference on Software Engineering and Knowledge Engineering (SEKE'13)* (Boston, USA, June 27-29, 2013). Knowledge Systems Institute, Skokie, Illinois, 2013, 455-460.

[9] Watson J. Psychology as the Behaviorist Views It. *Psychological Review*, 20, 2 (Feb. 1913), 158-177.

[10] Barabási, A.-L. The origin of bursts and heavy tails in human dynamics. *Nature*, 435, 7039 (May. 2005), 207-211.

[11] Dezsö, Z., Almaas, E., Lukács, A., Rácz, B., Szakadát, I. and Barabási, A.-L. Dynamics of information access on the web. *Physical Review E*, 73, 6 (Jun. 2006), 066132.

[12] Baek, S., Kim, T. and Kim, B. Testing a priority-based queue model with Linux command histories. *Physica A*, 387, 14 (Jun. 2008), 3660-3668.

[13] Baysal, O., Holmes, R. and Godfrey M. Mining usage data and development artifacts. In *Proceedings of the 9th IEEE Working Conference of Mining Software Repositories (MSR'12)* (Zurich, Switzerland, June 2-3, 2012). IEEE Computer Society Press, New York, NY, 2012, 98-107.

[14] Choi, J., Moon, J., Hahn, J. and Kim, J. Herding in open source software development: an exploratory study. In *Proceedings of the 16th ACM Conference on Computer Supported Cooperative Work (CSCW'13)* (San Antonio, USA, February 23-27, 2013). ACM, New York, NY, 2013, 129-134.

[15] Hofmann, P. and Riehle, D. Estimating Commit Sizes Efficiently. In *Proceedings of the 5th IFIP WG 2.13 International Conference on Open Source Systems (OSS 2009)* (Skövde, Sweden, June 3-6, 2009). Springer-Verlag, Heidelberg, Baden-Württemberg, 2009, 105-115.

[16] Alali, A., Kagdi, H. and Maletic, J. What's a Typical Commit? A Characterization of Open Source Software Repositories. In *Proceedings of the 16th IEEE International Conference on Program Comprehension (ICPC'08)* (Amsterdam, The Netherlands, June 10-13, 2008). IEEE Computer Society Press, New York, NY, 2008, 182-191.

[17] Hindle, A., Germán, D. and Holt, R. What do large commits tell us?: a taxonomical study of large commits. In *Proceedings of the 2008 International Working Conference on Mining Software Repositories (MSR'08)* (Leipzig, Germany, May 10-11, 2008). ACM Press, New York, NY, 2008, 99-108.

[18] Clauset, A., Shalizi, C. and Newman, M. Power-law distributions in empirical data. *SIAM Review*, 51, 4 (Apr. 2009), 661-703.

[19] Kolassa, C., Riehle, D. and Salim, M. The Empirical Commit Frequency Distribution of Open Source Projects. In *Proceedings of the 9th International Symposium on Open Collaboration (OpenSym'13)* (Hong Kong, China, Aug. 5-7, 2013). ACM Press, New York, NY, 2013, Article No. 0303.

[20] Singh, P. The small-world effect: The influence of macro-level properties of developer collaboration networks on open-source project success. *ACM Transactions on Software Engineering and Methodology*, 20, 2 (Mar. 2010), Article No. 6.

[21] Ma, Y., He, K., Li, B., Liu, J. and Zhou, X. A Hybrid Set of Complexity Metrics for Large-Scale Object-Oriented Software Systems. *Journal of Computer Science and Technology*, 25, 6 (Nov. 2010), 1184-1201.

[22] Chidamber, S. and Kemerer, C. A Metrics Suite for Object-Oriented Design. *IEEE Transactions on Software Engineering*, 20, 6 (Jun. 1994), 476-493.

[23] Olbrich, S., Cruzes, D. and Sjøberg, D. Are all code smells harmful? A study of God Classes and Brain Classes in the evolution of three open source systems. In *Proceedings of the 26th IEEE International Conference on Software Maintenance (ICSM'10)* (Timisoara, Romania, Sep. 12-18, 2010). IEEE Computer Society Press, New York, NY, 2010, 1-10.

[24] Moin, A. and Khansari, M. Bug Localization Using Revision Log Analysis and Open Bug Repository Text Categorization. In *Proceedings of the 6th International IFIP WG 2.13 Conference on Open Source Systems (OSS'10)* (Notre Dame, USA, May 30 - June 2, 2010). Springer-Verlag, Heidelberg, Baden-Württemberg, 188-199.

[25] http://en.wikipedia.org/wiki/Committer

[26] Eyolfson, J., Tan, L. and Lam, P. Correlations between Bugginess and Time-Based Commit Characteristics. *Empirical Software Engineering*, published online in Feb. (DOI: 10.1007/s10664-013-9245-0), 2013.